\documentclass[11pt]{article}
\usepackage{amsthm,amsmath,amssymb,color}
\usepackage{paralist}
\usepackage{lscape}
\usepackage{cite}

\newcommand{\ds}{\displaystyle}

\newcommand{\ord}{{\rm {ord}}}
\newcommand{\hdual}{{\bot_h}}

\newcommand{\floor}[1]{\left\lfloor {#1}\right\rfloor}
\newcommand{\ceil}[1]{\left\lceil {#1}\right\rceil}

\newtheorem{theorem}{Theorem}
\newtheorem{lemma}[theorem]{Lemma}

\newtheorem{corollary}[theorem]{Corollary}

\newcommand{\B}{{\mathcal{BCH}}}

\newcommand{\F}{\mathbf{F}}
\newcommand{\Z}{\mathbf{Z}}

\makeatletter
\renewcommand\section{\@startsection {section}{1}{\z@}%
                                   {-3.5ex \@plus -1ex \@minus -.2ex}%
                                   {2.3ex \@plus.2ex}%
                                  {\normalfont\Large\bfseries}}
\makeatother

\begin{document}

\title{\textbf{On Quantum and Classical BCH Codes}}
\author{Salah~A.~Aly, Andreas Klappenecker\footnote{Corresponding author, 
\texttt{klappi<at-sign>cs.tamu.edu}}, Pradeep
Kiran Sarvepalli\\
Department of Computer Science\\
Texas A\&M University\\
College Station, TX 77843-3112, USA\\[1ex]
} \maketitle
\begin{abstract}
\noindent 
Classical BCH codes that contain their (Euclidean or Hermitian) dual
codes can be used to construct quantum stabilizer codes; this
correspondence studies the properties of such codes.  It is shown that
a BCH code of length $n$ can contain its dual code only if its
designed distance $\delta=O(\sqrt{n})$, and the converse is proved in
the case of narrow-sense codes. Furthermore, the dimension of
narrow-sense BCH codes with small design distance is completely
determined, and -- consequently -- the bounds on their minimum distance are
improved.  These results make it possible to determine the parameters
of quantum BCH codes in terms of their design parameters.

\smallskip

\noindent\textit{Keywords:} quantum codes, BCH codes, dimension, minimum
distance, dual codes
\end{abstract}

\section{Introduction}
The Bose-Chaudhuri-Hocquenghem (BCH)
codes~\cite{Bose60a,Bose60b,Gorenstein61,Hocquenghem59} are a
well-studied class of cyclic codes that have found numerous
applications in classical and more recently in 
quantum information processing. Recall
that a cyclic code of length $n$ over a finite field $\F_q$ with $q$
elements, and $\gcd(n,q)=1$, is called a BCH code with designed
distance $\delta$ if its generator polynomial is of the form
$$g(x)=\prod_{z\in Z} (x-\alpha^z), \qquad Z=C_b\cup \cdots \cup
C_{b+\delta-2},$$ where $C_x=\{ xq^k\bmod n \,|\, k\in \Z, k\ge 0\,\}$
denotes the $q$-ary cyclotomic coset of $x$ modulo~$n$, $\alpha$ is a
primitive element of $\F_{q^m}$, and $m=\ord_n(q)$ is the
multiplicative order of $q$ modulo $n$.  Such a code is called primitive
if $n=q^m-1$, and narrow-sense if $b=1$.

An attractive feature of a (narrow-sense) BCH code is that one can
derive many structural properties of the code from the knowledge of
the parameters $n$, $q$, and $\delta$ alone.  Perhaps the most
well-known facts are that such a code has minimum distance $d\ge
\delta$ and dimension $k\ge n-(\delta-1)\ord_n(q)$.  In this
correspondence, we will show that a necessary condition for a 
narrow-sense BCH code which contains its
Euclidean dual code is that its designed distance $\delta=O(qn^{1/2})$.
We also derive a sufficient condition for dual containing BCH codes.
Moreover, if the codes are primitive, these conditions are same. These
results allow us to derive families of quantum stabilizer codes. Along
the way, we find new results concerning the minimum distance and
dimension of classical BCH codes.

To put our results into context, we give a brief overview of related
work.  This correspondence was motivated by problems concerning
quantum BCH codes; specifically, our goal was to derive the parameters
of the quantum codes as a function of the design parameters.  Examples
of certain binary quantum BCH codes have been given by many authors,
see, for example,~\cite{calderbank98,grassl99b,grassl00,steane96}.
Steane \cite{steane99} gave a simple criterion to decide when a binary
narrow-sense primitive BCH code contains its dual, given the design
distance and the length of the code.  We generalize Steane's result in
various ways, in particular, to narrow-sense (not necessarily
primitive) BCH codes over arbitrary finite fields with respect to
Euclidean and Hermitian duality. These results allow one to derive
quantum BCH codes; however, it remains to
determine the dimension, purity, and minimum distance of such quantum
codes.

The dimension of a classical BCH code can be bounded by many different
standard methods, see~\cite{berlekamp68,huffman03,macwilliams77} and
the references therein. An upper bound on the dimension was given by
Shparlinski~\cite{shparlinski88}, see
also~\cite[Chapter~17]{konyagin99}. More recently, the dimension of
primitive narrow-sense BCH codes of designed distance $\delta<
q^{\lceil m/2\rceil}+1$ was apparently determined by Yue and
Hu~\cite{yue96}, according to reference~\cite{yue00}.  We generalize
their result and determine the dimension of narrow-sense BCH codes
that are not necessarily primitive for a certain range of designed
distances. As desired, this result allows us to explicitly obtain the
dimension of the quantum codes without computation of cyclotomic
cosets.

The purity and minimum distance of a quantum BCH code depend on the
minimum distance and dual distance of the associated classical code.
In general, it is a difficult problem to determine the true minimum
distance of BCH codes, see~\cite{charpin98}.  A lower bound on the
dual distance can be given by the Carlitz-Uchiyama-type bounds when
the number of field elements is prime, see, for
example,~\cite[page~280]{macwilliams77} and \cite{stichtenoth94}. Many
authors have determined the true minimum distance of BCH codes in
special cases, see, for instance,~\cite{peterson72},\cite{yue00}.

This paper also extends our previous work on \textit{primitive}
narrow-sense BCH codes \cite{salah05}, simplifies some of the proofs
and generalizes many of the results to the nonprimitive case.

\medskip
\textit{Notation.} We denote the ring of integers by $\mathbf{Z}$ and
a finite field with $q$ elements by $\mathbf{F}_q$. We use the bracket
notation of Iverson and Knuth that associates to
$[\textit{statement}\,]$ the value 1 if \textit{statement} is true,
and 0 otherwise. For instance, we have $[k \text{ even}]= k-1\bmod 2$
and $[k \text{ odd}]= k\bmod 2$ for an integer $k$. The Euclidean dual
code $C^\perp$ of a code $C\subseteq \F_q^n$ is given by $C^\perp = \{
y\in \F_q^n \,|\, x\cdot y =0 \mbox{ for all } x \in C \},$ while the
Hermitian dual of $C\subseteq \F_{q^2}^n$ is defined as
$C^{\perp_h}=\{ y\in \F_{q^2}^n\,|\, y^q\cdot x = 0 \mbox{ for all } x
\in C\}$. We denote a narrow-sense BCH code of length $n$ over $\F_q$
with designed distance $\delta$ by $\B(n,q;\delta)$, and we omit the
parameter $q$ if the finite field is clear from the
context.

\section{Euclidean Dual Codes}
Recall that one can construct quantum stabilizer codes using classical
codes that contain their duals. In this section, our goal is to find
such classical codes. Steane showed that a primitive, narrow-sense,
binary BCH code of length $2^m-1$ contains its dual if and only if its
designed distance $\delta$ satisfies $\delta \leq 2^{\lceil
m/2\rceil}-1$, see~\cite{steane99}. We generalize this
result in various ways.

\begin{lemma}\label{th:selforthogonal}
Let $C$ be a cyclic code of length $n$ over the finite field $\F_q$
such that $\gcd(n,q)=1$, and let $Z$ be the defining set of $C$.  The
code $C$ contains its Euclidean dual code if and only if $Z\cap Z^{-1}
= \emptyset$, where $Z^{-1}$ denotes the set $Z^{-1}=\{-z\bmod n\mid z
\in Z \}$.
\end{lemma}
\begin{proof}
See \cite[Theorem~2]{grassl97}. See also~\cite[Theorem 4.4.11]{huffman03}.
\end{proof}

Let us first consider narrow-sense BCH codes of length $n$ such that
the multiplicative order of $q$ modulo $n$ equals 1; for example,
Reed-Solomon codes belong to this class of codes. We can avoid some
special cases in our subsequent arguments by treating this case
separately. Furthermore, the next lemma nicely illustrates the proof
technique that will be used throughout this section, so it can serve
as a warm-up exercise.

\begin{lemma} 
Suppose that $q$ is a power of a prime and $n$ is a positive integer
such that $q\equiv 1\bmod n$. We have $\B(n,q;\delta)^\perp\subseteq
\B(n,q;\delta)$ if and only if the designed distance $\delta$ is in
the range $2\le \delta\le \delta_{\max}=\lfloor (n+1)/2 \rfloor$.
\end{lemma}
\begin{proof}
The defining set $Z$ of $\B(n,q;\delta)$ is given by $Z=\{1,
\ldots,\delta-1\}$, since $q$ has multiplicative order 1 modulo $n$,
and therefore all cyclotomic cosets are singleton sets.  If $\B(n,q;\delta)^\perp\subseteq \B(n,q;\delta)$, then by Lemma~\ref{th:selforthogonal}, 
$Z\cap Z^{-1}=\emptyset$. If $x\in Z$, then $n-x\not\in Z$ and $n-x>x$; hence,
$\delta_{\max}\leq \lfloor (n+1)/2 \rfloor$. Conversely, if $\delta
\leq\lfloor (n+1)/2\rfloor$, then $\min Z^{-1} =\min \{n-1,\ldots,
n-\delta+1\} = n-\delta+1 \geq n-\lfloor (n+1)/2\rfloor +1 =\lceil (n+1)/2
 \rceil \geq \delta_{\max}$; hence, $Z\cap Z^{-1}=\emptyset$ and  Lemma~\ref{th:selforthogonal} implies that $\B(n,q;\delta)^\perp\subseteq \B(n,q;\delta)$.
\end{proof}

If the multiplicative order $m$ of $q$ modulo $n$ is larger than 1,
then the defining set of the code has a more intricate structure, so
proofs become more involved. The next theorem gives a sufficient 
condition on the designed distances 
for which the dual code of a narrow-sense BCH code is self-orthogonal. 

\begin{theorem}\label {th:sufficientEdual}
Suppose that $m=\ord_n(q)$. If the designed distance $\delta$
is in the range $2\leq \delta\leq \delta_{\max}=\floor{\kappa}$, where 
\begin{equation}\label{ddistbound}
\kappa = \frac{n}{q^{m}-1}
(q^{\lceil m/2\rceil}-1-(q-2)[m \textup{ odd}]), 
\end{equation} 
then $\B(n,q;\delta)^\perp \subseteq \B(n,q;\delta)$.
\end{theorem}
\begin{proof}
It suffices to show that $\B(n,q;\delta_{\max})^\perp\subseteq
\B(n,q;\delta_{\max})$ holds, since $\B(n,q;\delta)$ contains
$\B(n,q;\delta_{\max})$, and the claim follows from these two facts. 

Seeking a contradiction, we assume that $\B(n,q;\delta_{\max})$ does
not contain its dual.  Let $Z=C_1\cup \cdots \cup C_{\delta_{\max}-1}$
be the defining set of $\B(n,q;\delta_{\max})$.  By
Lemma~\ref{th:selforthogonal}, $Z\cap Z^{-1}\neq \emptyset$, which
means that there exist two elements $x,y\in
\{1,\ldots,\delta_{\max}-1\}$ such that $y\equiv -xq^j\bmod n$ for some
$j\in \{0,1,\ldots, m-1 \}$, where $m$ is the multiplicative order of
$q$ modulo $n$.  Since $\gcd(q,n)=1$ and $q^m\equiv 1 \bmod n$, we also
have $x\equiv -yq^{m-j}\bmod n$.  Thus, exchanging $x$ and $y$ if
necessary, we can even assume that $j$ is in the range $0 \leq j\leq
\lfloor m/2 \rfloor$.  It follows from (\ref{ddistbound}) that
$$1\leq  xq^j\le (\delta_{\max}-1)q^j\le 
\frac{n}{q^m-1}(q^m-q^j-q^j(q-2)[m\text{ odd}])-q^j<n, 
$$ for all $j$ in the range $0\le j\le \floor{m/2}$. Since $1\le
xq^j<n$ and $1\le y<n$, we can infer from $y\equiv -xq^j\bmod n$
that $y=n-xq^j$. But this implies 
$$
\begin{array}{lcl}
y &\ge&\ds n-xq^{\floor{m/2}} \\
&\ge& \ds n-\frac{n}{q^m-1}(q^m-q^{\floor{m/2}}
-q^{\floor{m/2}}(q-2)[m\text{ odd}])+q^{\floor{m/2}}\\
&=& \ds \frac{n}{q^m-1}( q^{\floor{m/2}}-1+q^{\floor{m/2}}(q-2)[m \text{ odd}]  )+q^{\floor{m/2}}\\
&\ge& \delta_{\max}\, , 
\end{array}
$$
contradicting the fact that $y<\delta_{\max}$.
\end{proof}

Now we will derive a necessary condition on the design distance of narrow-sense, nonprimitive BCH codes that contain their duals.

\begin{theorem}\label{th:necessaryEdual}
Suppose that $m=\ord_n(q)$. If the designed distance $\delta$ 
exceeds $\delta \geq \delta_{\max}=\floor{qn^{1/2}}$, 
then $\B(n,q;\delta)^\perp \not\subseteq \B(n,q;\delta)$.
\end{theorem}
\begin{proof}
Let $n=n_0+n_1q+\cdots+n_{d-1}q^{d-1}$, where $0\leq n_i\leq q-1$
and the defining set $Z=\{1,\ldots, \lfloor qn^{1/2}\rfloor \}$.
We will show that $Z\cap Z^{-1}\neq \emptyset$.
Let, 
\begin{eqnarray*}
s&=&\sum_{i=\lfloor d/2 \rfloor}^{ d-1}n_{i}q^{i-\lfloor d/2\rfloor},\\
s& \leq & (q-1)\sum_{i=\lfloor d/2\rfloor}^{ d-1} q^{i-\lfloor d/2\rfloor} =q^{\lceil d/2\rceil}-1 < q^{\lceil d/2\rceil}.
\end{eqnarray*}
Since $q^{d-1}<n< q^d$, we have $q^{(d+1)/2}< qn^{1/2}< q^{(d+2)/2}$. If
$d$ is even then $\lceil d/2\rceil < (d+1)/2$ and if $d$ is odd, then
$\lceil d/2 \rceil \leq (d+1)/2$. Hence we have $s< q^{\lceil d/2\rceil} \leq
q^{(d+1)/2}< qn^{1/2}$. Therefore $s\in Z$. Now consider,
\begin{eqnarray*}
s'=n-sq^{\lfloor d/2 \rfloor} &=& \sum_{i=0}^{d-1}n_iq^i-q^{\floor{ d/2}}
\sum_{i=\lfloor d/2\rfloor}^{ d-1}n_{i}q^{i-\lfloor d/2\rfloor},\\
&=&\sum_{i=0}^{\lfloor d/2 -1\rfloor}n_i q^i < q^{\floor{ d/2}} < q^{(d+1)/2}<qn^{1/2}.
\end{eqnarray*}
Hence $s'\in Z$ and by definition $s'\in Z^{-1}$, which implies $Z\cap Z^{-1}\neq\emptyset$; by Lemma~\ref{th:selforthogonal} it follows that $\B(n,q;\delta)^\perp\not\subseteq \B(n,q;\delta)$.
\end{proof}

The condition we just derived can be strengthened under some restrictions. 
Especially, if the constant $\kappa$ in equation~(\ref{ddistbound})is integral, then we can derive a necessary and sufficient 
condition as shown below:

\begin{theorem}\label{th:nonprimitiveduals2}
We keep the notation of Theorem~\ref{th:necessaryEdual}. 
Suppose that $\kappa$ is integral, and that $m\ge 2$. 
We have $\B(n,q;\delta)^\perp \subseteq \B(n,q;\delta)$ if and only if 
the designed 
distance $\delta$ is in the range $2\leq \delta\leq \delta_{\max}=\kappa$. 
\end{theorem}
\begin{proof}
Suppose that $\B(n,q;\delta)^\perp \subseteq
\B(n,q;\delta)$. Seeking a contradiction, we assume that 
$\delta>\delta_{\max}$; thus, $\delta_{\max}$ is
contained in the defining set $Z$ of $\B(n,q;\delta)$. 
If $m$ is even, then 
$$ -\delta_{\max}q^{\floor{m/2}}\equiv
-\frac{nq^{\floor{m/2}}}{q^{\floor{m/2}}+1}\equiv
-n+\frac{n}{q^{\floor{m/2}}+1}\equiv \delta_{\max}\pmod{n},
$$ hence, $\delta_{\max}\in Z\cap Z^{-1}\neq \emptyset$. If $m$ is
odd, then
$$\begin{array}{lcl}
-\delta_{\max}q^{\floor{m/2}}&\equiv&
-n(q^{m}-q^{\ceil{m/2}}+q^{\floor{m/2}})/(q^{m}-1)\\ 
&\equiv&
n(q^{\ceil{m/2}}-q^{\floor{m/2}}-1)/(q^m-1) \equiv s \pmod{n}.  
  \end{array}
$$ 
By definition, $s\in Z^{-1}$; furthermore,  $s<\delta_{\max}$, 
so $s\in Z\cap
Z^{-1}\neq \emptyset$. 
In both cases, $m$ even and odd, we found that $Z\cap Z^{-1}$ is not empty, so
$\B(n,q;\delta)$ cannot contain its Euclidean dual code,
contradiction.
The converse follows from Theorem~\ref{th:sufficientEdual}. 
\end{proof}
As a consequence of Theorem~\ref{th:nonprimitiveduals2} we have the following test for primitive narrow-sense BCH codes that contain their duals.
\begin{corollary}
A primitive narrow-sense BCH code of length $n=q^m-1$, $m\ge 2$, over
the finite field\/ $\F_q$ contains its Euclidean dual code if and only
if its designed distance $\delta$ satisfies
$$2\le \delta \leq
\delta_{\max}=q^{\lceil m/2\rceil}-1-(q-2)[m \textup{ odd}].$$
\end{corollary}

We observe that a narrow-sense BCH code containing its Euclidean dual
code must have a small designed distance ($\delta=O(\sqrt{n}))$, when
the multiplicative order of $q$ modulo $n$ is greater than one.  This
raises the question whether one can allow larger designed distances by
considering non-narrow-sense BCH codes. Our next result shows that
this is not possible, at least in the case of primitive codes.

\begin{theorem}\label {th:nonnarrowEdual}
Let $C$ be a primitive (not necessarily narrow-sense) BCH code of
length $n=q^m-1$ over $\F_q$ with designed distance $\delta$. If $m>1$ and
$\delta$ exceeds
$$
\delta_{\max}=\left\{ \begin{array}{ll}
q^{m/2}-1,& m \equiv 0\bmod 2,\\
2(q^{(m+1)/2}-q+1), & m \equiv 1 \bmod 2,
\end{array} \right. $$ then $C$
cannot contain its Euclidean dual.
\end{theorem}

\begin{proof}
Let the defining set of $C$ be $Z=C_{b}\cup C_{b+1}\cup \cdots \cup
C_{b+\delta-2}$. We will show that if $\delta > \delta_{\max}$ then
$Z\cap Z^{-1}\neq \emptyset$. If $0\in Z$, then $0\in Z^{-1}$, so
$Z\cap Z^{-1}\neq \emptyset$. Therefore, we can henceforth assume that
$0\not\in Z$, which implies $b\ge 1$ and $b+\delta-2<n$.
\begin{enumerate}
\item Suppose that $m$ is even; thus, $\delta_{\max}=q^{m/2}-1$.  If
$\delta > \delta_{\max}$ then the defining set $Z$ contains an element
of the form $s=\alpha \delta_{\max}$ for some integer
$\alpha$. However,
\begin{eqnarray*}
-sq^{m/2} &\equiv& -\alpha(q^{m/2}-1)q^{m/2} \equiv
 \alpha(q^{m/2}-1)\equiv s \pmod n.
\end{eqnarray*}
Hence, $s\in Z\cap Z^{-1}\neq \emptyset$. 
\item
Suppose that $m >1$ is odd; thus, $\delta_{\max}=2q^{(m+1)/2}-2q+2$. 
If $\delta> \delta_{\max}$ then there exists an integer $\alpha$ such that 
two multiples of $\delta'=\delta_{\max}/2$ are contained in the range 
$b\leq (\alpha-1) \delta' < \alpha\delta' \leq b+\delta-2$. 
Since $b\ge 1$ and $\alpha \delta'<n$, it follows that 
$2 \leq \alpha\leq q^{(m-1)/2}$.

The defining set $Z$ of the code contains the element
$s=\alpha\delta'$. The number $s'=\alpha(q^{(m+1)/2}-q^{(m-1)/2}-1)$
lies in the range $0\le s' \leq s$ and satisfies $-s q^{(m-1)/2}\equiv
s' \bmod n$, so $s'\in Z^{-1}$. 

Suppose that $b \leq s'$. Then $s'\in Z$, which implies $Z\cap
Z^{-1}\neq \emptyset$.

Suppose that $s'< b$. Since $b\leq (\alpha-1)\delta'$, we obtain
the inequality $s'<(\alpha-1)\delta'$; solving for $\alpha$ shows that
$\alpha\geq q$; thus, $q\leq \alpha\leq q^{(m-1)/2}$.  Let
$t'=(\alpha-1)(q^{(m+1)/2}-1)+q^{(m-1)/2}-1$; it is easy to check that
$t'$ is in the range $(\alpha-1)\delta'\le t'\le \alpha \delta'$
when $\alpha\geq q$; thus, $t'\in Z$.  Further, let $t=s-(\alpha-q+1)$; 
since $t\geq
s-\delta'$, we have $t\in Z$ as well. Since $-tq^{(m-1)/2}\equiv
t'\bmod n$, we can conclude that $t'\in Z\cap Z^{-1}\neq \emptyset$.
\end{enumerate}
Therefore, we can conclude that if the designed distance of $C$ is greater than
$\delta_{\max}$, then $Z\cap Z^{-1}\neq \emptyset$, which proves the claim 
thanks to Lemma~\ref{th:selforthogonal}. 
\end{proof}

\section{Dimension and Minimum Distance}
While the results in the previous section are sufficient to tell us when
we can construct quantum BCH codes, they are still unsatisfactory
because we do not know the dimension of these codes. To this end, we
determine the dimension of narrow-sense BCH codes of length $n$ with
minimum distance $d=O(n^{1/2})$. It turns out that these results on
dimension also allow us to sharpen the estimates of the true distance
of some BCH codes. 

First, we make some simple observations about
cyclotomic cosets that are essential in our proof.
\begin{lemma}\label{th:bchnpcosetsize}
Let $n$ be a positive integer and $q$ be a power of a prime such that
$\gcd(n,q)=1$ and $q^{\lfloor m/2\rfloor} <n \leq q^m-1$, where
$m=\ord_n(q)$. The cyclotomic coset $C_x=\{ xq^j\bmod n \mid 0\le
j<m\}$ has cardinality $m$ for all $x$ in the range $1\leq x\leq
nq^{\lceil m/2\rceil}/(q^m-1).$
\end{lemma}
\begin{proof}
If $m=1$, then $|C_x|=1$ for all $x$ and the statement is trivially
true. Therefore, we can assume that $m>1$.  Seeking a contradiction,
we suppose that $|C_x|<m$, meaning that there exists a divisor $j$ of
$m$ such that $xq^{j}\equiv x \bmod n$, or, equivalently, that
$x(q^{j}-1)\equiv 0\bmod n$ holds.

Suppose that $m$ is even. The divisor $j$ of $m$ must be in the range $1\le
j\le m/2$. However, $x(q^{j}-1)
\leq nq^{m/2}(q^{m/2}-1)/(q^{m}-1) <n $; hence $x(q^j-1)\not\equiv 0 \bmod
n$, contradicting the assumption $|C_x|<m$. 

Suppose that $m$ is odd. The divisor $j$ of $m$ must be in the range
$1\le j\le m/3$. Since $q^{(m+1)/2}\le q^{2m/3}$ for $m\ge 3$, we have
$x(q^j-1)\le nq^{(m+1)/2}(q^{m/3}-1)/(q^m-1)\le
nq^{2m/3}(q^{m/3}-1)/(q^m-1)<n$. Therefore, $x(q^j-1)\not\equiv 0\bmod
n$, contradicting the assumption $|C_x|<m$.
\end{proof}

The following observation tells us when some cyclotomic cosets
are disjoint.
\begin{lemma}\label{th:npdisjointcosets}
Let $n\ge 1$ be an integer and $q$ be a power of a prime such that
$\gcd(n,q)=1$ and $q^{\lfloor m/2\rfloor} <n \leq q^m-1$, where
$m=\ord_n(q)$. If $x$ and $y$ are distinct integers in the range
$1\leq x,\, y\leq \min\{ \lfloor nq^{\lceil m/2\rceil}/(q^m-1)-1\rfloor, n-1\}$
such that $x,y\not\equiv 0 \bmod q$, then the $q$-ary 
cyclotomic cosets of $x$ and $y$ modulo $n$ are distinct. 
\end{lemma}
\begin{proof} 
If $m=1$, then clearly $C_x=\{ x\}$, $C_y=\{y \}$ and distinct $x,y$ implies
that $C_x$ and $C_y$ are disjoint. If $m>1$, then $x,y\leq \lfloor nq^{\lceil m/2\rceil}/(q^m-1)-1 \rfloor <n-1$. The set 
$S=\{ xq^j\bmod n, yq^j \bmod n\,|\, 0\leq j\leq \lfloor m/2\rfloor\}$
contains $2(\lfloor m/2\rfloor+1) \geq m+1$ elements, since 
$q^{\floor{m/2}}\times \lfloor nq^{\lceil m/2\rceil} /(q^m-1)-1\rfloor < n$ and, thus, no two elements are identified modulo $n$.  If we assume
that $C_x=C_y$, then the preceding observation would imply that
$|C_x|=|C_y|\geq |S| \geq m+1$, which is
impossible since the maximal size of a cyclotomic coset is $m$.
Hence, the cyclotomic cosets $C_x$ and $C_y$ must be disjoint.
\end{proof}

With these results in hand, we can now derive the dimension of
narrow-sense BCH codes. 
\begin{theorem}\label{th:bchnpdimension}
Let $q$ be a prime power and $\gcd(n,q)=1$ with $\ord_n(q)=m$. Then a
narrow-sense BCH code of length $q^{\lfloor m/2\rfloor} <n \leq q^m-1$
over $\F_q$ with designed distance $\delta$ in the range $2 \leq \delta
\le \min\{ \lfloor nq^{\lceil m/2 \rceil}/(q^m-1)\rfloor,n\} $ has dimension
\begin{equation}\label{eq:npdimension}
k=n-m\lceil (\delta-1)(1-1/q)\rceil.
\end{equation}
\end{theorem}
\begin{proof}
Let the defining set of $\B(n,q;\delta)$ be $Z=C_1\cup C_2\cdots \cup
C_{\delta-1}$; a union of at most $\delta -1$ consecutive cyclotomic
cosets. However, when $1\leq x\leq \delta-1$ is a multiple of $q$, then
$C_{x/q}=C_x$. Therefore, the number of cosets is reduced by
$\lfloor(\delta-1)/q \rfloor$. By Lemma~\ref{th:npdisjointcosets}, if $x,
y\not\equiv 0 \bmod q$ and $x\neq y$, then the cosets $C_x$ and $C_y$ are
disjoint. Thus, $Z$ is the union of $(\delta-1)-\lfloor
(\delta-1)/q\rfloor= \lceil (\delta-1)(1-1/q)\rceil$ distinct cyclotomic
cosets. By Lemma~\ref{th:bchnpcosetsize}, all these cosets have
cardinality~$m$. Therefore, the degree of the generator polynomial is
$m\lceil (\delta-1)(1-1/q)\rceil$, which proves our claim about the
dimension of the code.
\end{proof}

As a consequence of the dimension result, we can tighten the bounds on
the minimum distance of narrow-sense BCH codes generalizing a
result due to Farr, see~\cite[p.~259]{macwilliams77}.

\begin{corollary}
A $\B(n,q;\delta)$ code
\begin{compactenum}[i)]
\item with length in the range $q^{\lfloor m/2\rfloor} <n \leq q^m-1$, 
$m=\ord_n(q)$, 
\item and designed distance in the range 
$2 \leq \delta \le \min\{ \lfloor nq^{\lceil m/2 \rceil}/(q^m-1)\rfloor,n\} $ 
\item such that 
\begin{eqnarray}\label{eqa3}
\sum_{i=0}^{\lfloor (\delta+1)/2\rfloor} \binom{n}{i} (q-1)^i
>q^{m\lceil (\delta-1)(1-1/q)\rceil},
\end{eqnarray}
\end{compactenum}
has minimum distance $d= \delta$ or $\delta+1$; if $\delta\equiv 0\bmod q$, then $d=\delta+1$.
\end{corollary}
\begin{proof}
Seeking a contradiction, we assume that the minimum distance~$d$ of the
code satisfies $d \geq \delta+2$. We know from
Theorem~\ref{th:bchnpdimension} that the dimension of the code is
$k=n-m\lceil (\delta-1)(1-1/q)\rceil.$ If we substitute this value of $k$
into the sphere-packing bound
$q^{k}   \sum_{i=0}^{\lfloor (d-1)/2\rfloor}
\binom{n}{i} (q-1)^i \leq q^n$, 
then we obtain
$$
\sum_{i=0}^{\lfloor (\delta+1)/2\rfloor}\binom{n}{i}(q-1)^i\le
\sum_{i=0}^{\lfloor (d-1)/2\rfloor}\binom{n}{i}(q-1)^i\le q^{m\lceil
(\delta-1)(1-1/q)\rceil},
$$
but this contradicts condition~(\ref{eqa3}); hence,
$\delta\le d\le \delta+1$.

If $\delta\equiv 0\bmod q$, then the cyclotomic coset $C_\delta$ is
contained in the defining set $Z$ of the code because
$C_\delta=C_{\delta/q}$. Thus, the BCH bound implies that the minimum
distance must be at least $\delta+1$.
\end{proof}

We conclude this section with  a minor result on the dual distance of
BCH codes which will be needed later for determining the purity of quantum
codes. 
\begin{lemma}\label{th:Edualdist}
Suppose that $C$ is a narrow-sense BCH code of length $n$ over $\F_q$
with designed distance $2\leq \delta\le \delta_{\max}=\lfloor n(q^{\lceil
m/2\rceil}-1-(q-2)[m \textup{ odd}])/(q^m-1) \rfloor$, then the dual distance
$d^\perp \geq \delta_{\max} + 1$.
\end{lemma}
\begin{proof}
Let $N=\{0,1,\ldots,n-1 \}$ and $Z_{\delta}$ be the defining set of
$C$. We know that $Z_{\delta_{\max}}\supseteq Z_{\delta}\supset
\{1,\ldots,\delta-1 \}$.  Therefore $N\setminus Z_{\delta_{\max}}
\subseteq N\setminus Z_{\delta}$.  Further, we know that $Z\cap
Z^{-1}=\emptyset$ if $2\leq \delta\leq \delta_{\max }$ from
Lemma~\ref{th:selforthogonal} and
Theorem~\ref{th:sufficientEdual}. Therefore,
$Z^{-1}_{\delta_{\max}}\subseteq N\setminus Z_{\delta_{\max}}\subseteq
N\setminus Z_{\delta}$.

Let $T_{\delta}$ be the defining set of the dual code. Then
$T_{\delta}=(N\setminus Z_{\delta})^{-1} \supseteq
Z_{\delta_{\max}}$. Moreover $\{0\}\in N\setminus Z_{\delta}$ and
therefore $T_{\delta}$. Thus there are at least $\delta_{\max}$
consecutive roots in $T_{\delta}$. Thus the dual distance $d^\perp
\geq \delta_{\max}+1$.
\end{proof}

\section{Hermitian Dual Codes}
Suppose that $C$ is a linear code of length $n$ over $\F_{q^2}$.
Recall that its Hermitian dual code is defined by
$C^{\perp_h}=\{ y\in \F_{q^2}^n\,|\, y^q\cdot x = 0 \mbox{ for all } x
\in C\}$, where $y^q=(y_1^q,\dots,y_n^q)$ denotes the conjugate of the
vector $y=(y_1,\dots,y_n)$.

\begin{lemma}\label{th:hermitian}
Assume that $\gcd(n,q)=1$. A cyclic code of length $n$ over
$\F_{q^2}$ with defining set $Z$ contains its Hermitian dual code if
and only if $Z\cap Z^{-q} = \emptyset$, where $Z^{-q}=\{-qz \bmod n
\mid z \in Z \}$.
\end{lemma}
\begin{proof}
Let $N=\{0,1,\dots,n-1\}$. If $g(x)=\prod_{z\in Z} (x-\alpha^z)$ is the
generator polynomial of a cyclic code $C$, then $h^\dagger(x)=\prod_{z\in
N\setminus Z} (x-\alpha^{-qz})$ is the generator polynomial of
$C^{\perp_h}$.  Thus, $C^{\perp_h}\subseteq C$ if and only if $g(x)$
divides $h^\dagger(x)$. The latter condition is equivalent to $Z\subseteq
\{ -qz\,|\, z\in N\setminus Z\}$, which can also be expressed as $Z\cap
Z^{-q}=\emptyset$.
\end{proof}

Now similar to Theorem~\ref {th:sufficientEdual} we will derive a 
sufficient condition for BCH codes that contain their Hermitian duals.

\begin{theorem}\label {th:sufficientHdual}
Suppose that $m=\ord_n(q^2)$. If the designed distance $\delta$ satisfies
$$2\leq \delta \leq \delta_{\max}=\left\lfloor \frac{n}{q^{2m}-1}
(q^{ m+[\textup{m  even}] }-1-(q^2-2)[m \textup{ even}])\right\rfloor,$$
then $\B(n,q^2;\delta)^{\perp_h} \subseteq
\B(n,q^2;\delta).$
\end{theorem}
\begin{proof}
Since $\B(n,q^2;\delta)$ contains $\B(n,q^2;\delta_{max})$, it
suffices to show that $\B(n,q^2;\delta_{max})^{\perp_h} \subseteq
\B(n,q^2;\delta_{max})$ holds.

Seeking a contradiction, we assume that $\B(n,q^2;\delta_{max})$ does not
contain its dual. Let $Z=C_1 \cup C_2\cup \dots \cup C_{\delta_{\max}-1}$
be the defining set of $\B(n,q^2;\delta_{max} )$. By
Lemma~\ref{th:hermitian}, $Z\cap Z^{-q}\neq \emptyset$, which means that
there exist two elements $x,y \in \{1,...,\delta_{max}-1\}$ such that
$y=-xq^{2j+1} \bmod n$ for some $j \in \{0,1,...,m-1\}$, where
$m=\ord_n(q)$. Since $\gcd(q,n)=1$ and $q^{2m} \equiv 1 \bmod n$, we
also have $y \equiv - xq^{2m-2j-1} \bmod n$, so we can assume without loss
of generality that $j$ lies in the range $0 \leq j \leq \lfloor (m-1)/2
\rfloor$. It follows that
\begin{eqnarray*}
xq^{2j+1} &\leq& (\delta_{max}-1)q^{2j+1} \\ &=&
\frac{nq^{2j+1}}{q^{2m}-1} (q^{ m+[\textup{m  even}] }-1-(q^2-2)[m
\textup{ even}]) -q^{2j+1} \\&<& n
\end{eqnarray*}
holds  for all $j$ in the range $0 \leq j \leq \lfloor (m-1)/2\rfloor$.

Since $1 \leq xq^{2j+1} < n$, the congruence $y \equiv -xq^{2j+1}
\bmod n$ implies that $y=n-xq^{2j+1}$. Therefore, $y\ge
n-(\delta_{\max}-1)q^{2\floor{(m-1)/2}+1}$, which is equivalent to 
$$
y\geq n -  \frac{n q^{2\lfloor (m-1)/2\rfloor+1}}{q^{2m}-1} 
(q^{ m+[\textup{m even}]}-1-(q^2-2)[m \textup{ even}]) 
+q^{2\floor{(m-1)/2}+1}.
$$  
If $m$ is odd, this yields 
\begin{eqnarray*}
y&\geq& n -  
\frac{n q^m}{q^{2m}-1} (q^{ m }-1)+q^m
= \frac{n}{q^{2m-1}}(q^{m}-1)+q^m\ge \delta_{\max}\, .
\end{eqnarray*}
Similarly, if $m$ is even, then 
\begin{eqnarray*}
   y&\geq&\frac{n}{q^{2m}-1}(q^{m+1}-q^{m-1}-1)+q^{m-1} \ge \delta_{\max}. 
\end{eqnarray*}
Both cases contradict the assumption $0\le y<\delta_{\max}$. Therefore, we
can conclude that $\B(n,q;\delta_{\max})$ contains its Hermitian
dual code.
\end{proof}
Arguing as in Theorem~\ref{th:necessaryEdual} we can show that a BCH code 
must have its designed distance $\delta=O(q^2n^{1/2})$ if it contains its 
Hermitian dual. As the arguments are very similar we illustrate it for a
simpler case as shown below:

\begin{lemma}
Let $C\subseteq \F_{q^2}^n$ be a nonnarrow-sense, nonprimitive BCH
code of length $n\equiv 0\bmod q^{m}+1$, where $m=\ord_n(q^2)$. If its
design distance $\delta \geq \delta_{\max}=n/(q^{m}+1)$, then $C$
cannot contain its Hermitian dual.
\end{lemma}
\begin{proof}
The defining set $Z=C_b\cup \ldots \cup C_{b+\delta-2}$ contains 
$\{ b,\ldots,b+\delta-2\}$. If
$\delta>\delta_{\max}=n/(q^{m}+1)$, then there exists an element
$s=\alpha\delta_{\max}\in Z$ for some positive integer $\alpha$.  Then
$-qs(q^2)^{(m-1)/2}\equiv-\alpha nq^{m}/(q^m+1) \equiv \alpha
n/(q^m+1)\equiv s \mod {n}$. Therefore, $Z\cap Z^{-q}\neq \emptyset$; hence, 
$C$ cannot contain its Hermitian dual code.
\end{proof}
Finally, we conclude this section on Hermitian duals by proving as in the 
Euclidean case nonnarrow-sense BCH codes that contain their Hermitian duals
cannot have too large design distances. 

\begin{theorem}\label{th:nonnarrowhdual}
Let $C \subseteq \F_{q^2}^n$ be a primitive (not necessarily
narrow-sense) BCH code of length $n=q^{2m}-1$, $m=\ord_n(q)$, and
designed distance $\delta$. If $\delta$ exceeds
$$\delta_{\max}=\left\{ \begin{array}{ll}
q^{m}-1 & \text{if\/ $m$ is odd},\\
2(q^{m+1}-q^2+1) & \text{if\/ $m\neq 2$ is even},
\end{array} \right. $$ then $C$ cannot contain its Hermitian dual code.
\end{theorem}
\begin{proof}
Suppose that the defining set of $C$ is given by $Z=C_b\cup \cdots
\cup C_{b+\delta-2}$, where $C_x=\{ xq^{2j}\bmod n\,|\, j\in \Z \}$,
and that $\delta>\delta_{\max}$. Seeking a contradiction, we assume
that $C^\hdual\subseteq C$, which means that $Z\cap Z^{-q}=\emptyset$.
It follows that $0\not\in Z$, for otherwise $0\in Z\cap Z^{-q}$;
therefore, $b\ge 1$ and $b+\delta-2<n$.

If $m$ is odd, then there exists an integer $\alpha$ such that $b\le
\alpha\delta_{\max}\le b+\delta-2$. We have
$-q\alpha\delta_{\max}q^{m-1} \equiv \alpha (1-q^m)q^m \equiv \alpha
(q^m-1) \equiv \alpha\delta_{\max}\bmod n$; thus,
$\alpha\delta_{\max}\in Z\cap Z^{-q}\neq \emptyset.$

If $m>2$ is even and $\delta> \delta_{\max}=2q^{m+1}-2q^2+2$,
then there exists an integer $\alpha$ such that 
two multiples of $\delta'=\delta_{\max}/2$ are contained in the range 
$b\leq (\alpha-1) \delta' < \alpha\delta' \leq b+\delta-2$. 
Since $b\ge 1$ and $\alpha \delta'<n$, it follows that 
$2 \leq \alpha\leq q^{m-1}$ (which holds only if $m>2$).

Clearly $s=\alpha\delta'\in Z$.
Let $s'\equiv -qsq^{m-2}\bmod n$, so $s'\in Z^{-q}$, then 
$1\leq s'=\alpha(q^{m+1}-q^{m-1}-1) \leq s$ for $m>2$.

Suppose that $b \leq s'$. Then $s'\in Z$, which implies $Z\cap
Z^{-q}\neq \emptyset$.

Suppose that $s'< b$. Since $b\leq (\alpha-1)\delta'$, we obtain
the inequality $s'<(\alpha-1)\delta'$; solving for $\alpha$ shows that
$\alpha\geq q^2$; thus, $q^2\leq \alpha\leq q^{m-1}$.  Let
$t'=(\alpha-1)(q^{m+1}-1)+q^{(m-1)/2}-1$; it is easy to check that
$t'$ is in the range $(\alpha-1)\delta'\le t'\le \alpha \delta'$
when $\alpha\geq q^2$; thus, $t'\in Z$.  Further, let $t=s-(\alpha-q^2+1)$; 
since $t\geq s-\delta'$, we have $t\in Z$ as well. Since $-qtq^{m-2}\equiv
t'\bmod n$, we can conclude that $t'\in Z\cap Z^{-q}\neq \emptyset$.
Hence, by Lemma~\ref{th:hermitian} we conclude that $C$ cannot contain its
Hermitian dual if its design distance exceeds $\delta_{\max}$
\end{proof}

\section{Families of Quantum BCH Codes}
In this section we shall study the construction of (nonbinary) quantum BCH codes.
Calderbank, Shor, Rains and Sloane outlined  the construction of binary quantum BCH codes  in \cite{calderbank98}. Grassl, Beth and Pellizari developed the theory further by  formulating a nice condition for determining which BCH codes can be used  for constructing quantum codes \cite{grassl97,grassl99b}. 
The dimension and the purity of the quantum codes constructed were 
determined by numerical computations. Steane simplified it further for the 
special case of binary narrow-sense primitive BCH codes \cite{steane99}
and gave a very simple criterion based on the design distance alone. 
Very little was done with respect to the nonprimitive and 
nonbinary quantum BCH codes. 

In this section we show how the results we have developed in the previous sections
help us to generalize the previous work on quantum codes and give very simple 
conditions based on design distance alone. Further, we give precisely the dimension
and tighten results on the purity of the quantum codes.  But, first we review the
methods of constructing quantum codes from classical codes. 

\begin{lemma}[Quantum Code Constructions] \label{th:qconst} \ \par
\begin{compactenum}[a)]
\item If there exists classical linear codes $C_1\subseteq C_2 \subseteq \F_q^n$,
then there exists an $[[n,k_2-k_1,d]]_q$ quantum code where 
$d=\min\{ (C_2\setminus C_1)\cup (C_1^\perp\setminus C_2^\perp)\}$.

\item If there exists a classical linear $[n,k,d]_q$ code $C$ such that
$C^\perp\subseteq C$, then there exists an $[[n,2k-n,\ge d]]_q$
stabilizer code that is pure to $d$. If the minimum distance of
$C^\perp$ exceeds $d$, then the stabilizer code is pure and has
minimum distance $d$.

\item If there exists a classical linear $[n,k,d]_{q^2}$ code $D$ such
that $D^\hdual\subseteq D$, then there exists an $[[n,2k-n,\ge
d]]_{q^2}$ stabilizer code that is pure to $d$. If the minimum
distance $d^\hdual$ of $D^\hdual$ exceeds $d$, then the stabilizer
code is pure and has minimum distance $d$.
\end{compactenum}
\end{lemma}
\begin{proof}
Part b) is a special case of a) which is commonly
referred to as the CSS construction and part c) is the
Hermitian code construction, see, for instance,~\cite{ketkar05}
for the proofs.
\end{proof}

\begin{theorem}\label{sh:nested}
Let $m=\ord_n(q)\geq 2$, where $q$ is a power of a prime and $\delta_1,\delta_2$ are integers such that $2\le \delta_1 < \delta_2\le \delta_{\max}$ where 
$$\delta_{\max} = \frac{n}{q^m-1}(q^{\lceil m/2\rceil}-1-(q-2)[m \textup{ odd}]),$$
then there exists a quantum code with parameters
$$[[n,m(\delta_2-\delta_1-\lfloor(\delta_2-1)/q\rfloor+\lfloor(\delta_1-1)/q \rfloor),\geq \delta_1]]_q$$
pure to $\delta_2$.
\end{theorem}
\begin{proof}
By Theorem~\ref{th:bchnpdimension}, there exist BCH codes $\B(n,q;\delta_i)$ with the parameters $[n,n-m(\delta_i-1)+m\lfloor(\delta_i-1)/q\rfloor,\geq \delta_i]_q$ for 
$i\in \{ 1,2\}$. Further, $\B(n,q;\delta_2)\subset \B(n,q;\delta_1)$. Hence by the CSS construction there exists a quantum code with the parameters
 $$[[n,m(\delta_2-\delta_1
-\lfloor(\delta_2-1)/q\rfloor +\lfloor(\delta_1-1)/q\rfloor),\geq \delta_1]]_q.$$ The purity follows due to the fact that $\delta_2>\delta_1$ and Lemma~\ref{th:Edualdist} by which the dual distance of either BCH code is $\geq \delta_{\max}+1 >\delta_2$. 
\end{proof}
When the BCH codes contain their duals, then we can derive the following codes. 
Note that these cannot be obtained as a consequence of Theorem~\ref{sh:nested}.

\begin{theorem}\label{sh:euclid}
Let $m=\ord_n(q)$  where $q$ is a power of a prime and 
$2\le \delta\le \delta_{\max},$ with 
$$\delta_{\max}=\frac{n}{q^m-1}(q^{\lceil m/2\rceil}-1-(q-2)[m  \textup{ odd}]),$$ 
then there exists a quantum code with parameters
$$[[n,n-2m\lceil(\delta-1)(1-1/q)\rceil,\ge \delta]]_q$$ pure to $\delta_{\max}+1$
\end{theorem}
\begin{proof}
Theorems~\ref {th:sufficientEdual} and \ref{th:bchnpdimension} imply that there
exists a classical BCH code with parameters
$[n,n-m\lceil(\delta-1)(1-1/q)\rceil,\ge \delta]_q$ which
contains its dual code. By Lemma~\ref{th:qconst}~b) an $[n,k,d]_q$ code that contains its dual
code implies the existence of the quantum code with parameters
$[[n,2k-n,\ge d]]_q$. The purity follows from Lemma~\ref{th:Edualdist} by 
which the dual distance $\geq \delta_{\max}+1 > \delta$.
\end{proof}

Before we can construct quantum codes via the Hermitian construction, we will 
need the following lemma. 
\begin{lemma}\label{th:Hdualdist} 
Suppose that $C$ is a primitive, narrow-sense BCH code of length
$n=q^{2m}-1$ over $\F_{q^2}$ with designed distance $2\leq \delta\le
\delta_{\max}=q^{m+[\text{$m$ even}]}-1-(q^2-2)[m \textup{ even}])$, 
then the dual distance $d^\perp \geq
\delta_{\max} + 1$.
\end{lemma}
\begin{proof}
The proof is analogous to the one of Lemma~\ref{th:Edualdist}; just
keep in mind that the defining set $Z_\delta$ is invariant under
multiplication by $q^2$ modulo $n$.
\end{proof}

\begin{theorem}\label{sh:hermite}
Let $m=\ord_n(q^2) \geq 2$ where $q$ is a power of a prime
and  $2\le \delta \le \delta_{\max}=\lfloor n(q^{m}-1)/(q^{2m}-1)\rfloor$, then
there exists a quantum code with parameters
$$ [[n, n-2m\lceil(\delta-1)(1-1/q^2)\rceil ,\ge \delta]]_q$$
that is pure up to $\delta_{\max}+1$.
\end{theorem}
\begin{proof}
It follows from Theorems~\ref{th:bchnpdimension} and~\ref {th:sufficientHdual} that
there exists a primitive, narrow-sense $[n,n-1-m\lceil
(\delta-1)(1-1/q^2)\rceil,\ge\delta]_{q^2}$ BCH code that contains its
Hermitian dual code.  By Lemma~\ref{th:qconst}~c) a classical $[n,k,d]_{q^2}$ code
that contains its Hermitian dual code implies the existence of an
$[[n,2k-n,\ge d]]_q$ quantum code. By Lemma~\ref{th:Hdualdist} the quantum code is
pure to $\delta_{\max}+1$.
\end{proof}
In the above theorem, quantum codes can also be constructed when the design 
distance exceeds the given value of $\delta_{\max}$, however we do not have exact
knowledge of the dimension in all those cases, hence we have not included them to keep
the theorem precise.

These are not the only possible families of quantum codes that can be derived from
BCH codes. As pointed out in \cite{grassl99b}, we can expand 
BCH codes over $\F_{q^l}$ to get codes over $\F_q$. Once again the dimension
and duality results of BCH codes makes it very easy to specify such codes. We will
just give one example in the Euclidean case. Similar results can be derived for the
Hermitian case. 

\begin{theorem}\label{sh:euclidexpansion}
Let $m=\ord_n(q^l)$  where $q$ is a power of a prime and 
$2\le \delta\le \delta_{\max},$ with 
$$\delta_{\max}=\frac{n}{q^{lm}-1}(q^{l\lceil m/2\rceil}-1-(q^l-2)[m 
\textup{ odd}]),$$ 
then there exists a quantum code with parameters
$$[[ln,ln-2lm\lceil(\delta-1)(1-1/q)\rceil,\ge \delta]]_q$$
that is pure up to $\delta$.
\end{theorem}
\begin{proof}
By Theorem~\ref{sh:euclid} there
exists a quantum BCH code with parameters
$[[n,n-2m\lceil(\delta-1)(1-1/q)\rceil,\ge \delta]]_{q^l}$.
An $[[n,k,d]]_{q^l}$ quantum code
implies the existence of the quantum code with parameters
$[[ln,lk,\ge d]]_q$ by \cite[Lemma~76]{ketkar05} and the code follows.
\end{proof}


\section{Conclusions}
In this paper  we have identified the classes of 
BCH codes that contain their Euclidean (Hermitian) duals by a careful 
analysis of the cyclotomic cosets. In the process
we have been able to shed more light on the structure of
dual containing BCH codes. We were able to 
derive a formula for the dimension of narrow-sense BCH
codes when the designed distance is small. 
These results allowed us to identify easily which classical BCH codes can 
be used for construct quantum codes. Further, the parameters
of these quantum codes are easily specified in terms of the design
distance.

\paragraph{Acknowledgments.\kern-4.38275pt}
This research was supported by NSF CAREER award CCF~0347310,
NSF grant CCF~0218582, and Texas A\&M TITF initiative.




\end{document}